\def \myplotone#1 {\centerline{#1}}
\def \myplotfiddle#1 {\centerline{#1}}

\def \vectheta {{\vec \theta}}

\def \etal {{\it et al.\/}}
\def \Sigmacrit {\Sigma_{\rm crit}}

\documentstyle[aaspp]{article}
\eqsecnum

\begin{document}

\title{A Method for Weak Lensing Observations}
\author{Nick Kaiser\altaffilmark{1}, Gordon Squires\altaffilmark{2},
Tom Broadhurst\altaffilmark{3}}
\altaffiltext{1}{Canadian Institute for Advanced Research and\\
Canadian Institute for Theoretical Astrophysics, University of Toronto\\
60 St.\ George St., Toronto, Ontario, M5S 1A7\\
kaiser@cita.utoronto.ca}
\altaffiltext{2}{Physics Department, University of Toronto\\
60 St.\ George St., Toronto, Ontario, M5S 1A7\\
squires@cita.utoronto.ca}
\altaffiltext{3}{Astronomy Department, Johns Hopkins University\\
Baltimore MD 21218\\
tjb@skysrv.pha.jhu.edu}

\begin{abstract}
We develop and test a method for measuring the gravitational lensing induced
distortion of faint background galaxies.  We first describe how we locate the
galaxies and measure a 2-component `polarisation' or ellipticity statistic
$e_\alpha$ whose expectation value should be proportional to the gravitational
shear $\gamma_\alpha$.  We then show that an anisotropic instrumental psf
perturbs the polarisation by $\delta e_\alpha = P^s_{\alpha\beta} p_\beta$,
where $p_\alpha$ is a measure of the psf anisotropy and $P^s_{\alpha\beta}$
is the `linearised smear polarisability tensor'.  By estimating
$P^s_{\alpha\beta}$ for each object we can determine $p_\alpha$ from the
foreground stars and apply a correction $-P^s_{\alpha\beta}p_\beta$ to the
galaxies.  We test this procedure using deep high-resolution images from HST
which are smeared with an anisotropic psf and then have noise added to simulate
ground-based observations.  We find that the procedure works very well.  A
similar analysis yields a linear shear polarisability tensor
$P^\gamma_{\alpha\beta}$ which describes the response to a gravitational shear.
This calibrates the polarisation-shear relation, but only for galaxies which
are well resolved.  To empirically calibrate the effect of seeing on the
smaller
galaxies we artificially stretch HST images to simulate lensing and then
degrade
them as before.  These experiments provide a rigorous and exacting test of the
method under realistic conditions. They show that it is possible to remove the
effect of instrumental psf anisotropy, and that the method provides an
efficient
and quantitative measurement of the gravitational shear.
\end{abstract}


\section{Introduction}
Gravitational lensing of faint background galaxies
provides a powerful probe of the mass distribution in and around
clusters (Tyson, Valdes and Wenk, 1990; Miralda-Escude, 1991a;
Kochanek, 1991; Bonnet \etal, 1994;
Bonnet and Mellier, 1994; Mellier, \etal, 1994; Smail  \etal 1994a; Smail
\etal, 1994b;
Fahlman \etal, 1994; Dahle \etal; 1994;
Schneider and Seitz, 1994; Seitz and Schneider, 1994; Schramm and Kayser, 1994;
Broadhurst \etal, 1994; Fort and Mellier, 1994;
Kaiser \etal, 1994, Tyson, 1994)
and potentially of large-scale structure
(Blandford \etal, 1991; Miralda-Escude, 1991b; Kaiser, 1992; Mould \etal, 1994;
Gould and Villumsen, 1994).
Lensing will both amplify
and distort the images of background galaxies.  Here we will restrict attention
to the statistical
anisotropy of the background galaxies caused by the tidal shearing
of the light rays.

With perfect seeing, the effect of a gravitational lens is a simple
Lagrangian mapping
of the surface brightness pattern
\begin{equation}	\label{eq:mapping}
f'(\theta_i) = f(\psi_{ij} \theta_j)
\end{equation}
where $f'$ is the observed surface brightness and $f$ that which would have
been
observed in the absence of lensing and
where angles are measured relative to some fiducial point on the image
(e.g.~the centre).
In the weak lensing regime --- the main subject of this paper --- the
image shear tensor $\psi_{ij}$ is close to the unit matrix, and
$\psi_{ij} - \delta_{ij}$ is just
an integral along the line of sight of the transverse components of the
tidal field (e.g.~Kaiser, 1992). For the special case of
a planar lens, $\psi_{ij} = \delta_{ij} - \phi_{ij}$, where the dimensionless
{\sl surface potential\/} $\phi$ is related to the
projected mass density by Poisson's
equation in 2-dimensions: $\nabla^2 \phi = 2 \Sigma/ \Sigmacrit$.
For an Einstein de Sitter universe the inverse critical surface density
is $\Sigmacrit^{-1} \equiv 4 \pi a_l w_l \beta$ where $w_l$ is the
comoving distance to the lens, $w_l = 1 - (1+z_l)^{-1/2}$ and
$a_l$ is the scale factor at $z_l$, and the factor $\beta \equiv \max(0, 1 -
w_l / w_s)$
gives the distortion strength as a function of the source
distance $w_s$ ($\beta \rightarrow 1$ for $w_s \gg w_l$).

For a spatially constant $\psi_{ij}$
the mapping (eq.~\ref{eq:mapping}) describes a simple anisotropic dilation of
the
images much as though the galaxies were painted on a rubber sheet
which is then stretched.  The ratio of the stretch factors along the
axes in the diagonal frame (for an
intrinsically circular object this is just ratio of the major and
minor axes) is equal to
the ratio of the eigenvalues
of the tensor $\psi_{ij}$. In reality we will see
a superposition of source planes at different redshifts which will
have been distorted by different amounts (though for cluster lenses
at $z \lesssim 0.2$ the distribution of $\beta$ values is actually
quite narrow, see \S\ref{sec:shearmeasurement}, and the single sheet
approximation is quite
good). The goal of the research described here was to develop techniques to
measure the mean anisotropy
strength and orientation $\varphi$
for the galaxies on some patch of sky; in the weak distortion regime
($\phi_{ij}\ll 1$) the distortion is proportional to the {\sl shear} $\gamma_i
\equiv
\gamma \{\cos 2 \varphi, \sin 2 \varphi \}$, where
$\gamma \equiv
(\lambda_1 - \lambda_2)/2$, with $\lambda_1,\lambda_2$ the eigenvalues
of $\phi_{ij}$.  How the surface density $\Sigma$ may be reconstructed from
measurement of $\gamma$ has been discussed in more detail
elsewhere (KS93; Kaiser \etal, 94,
Schneider and Seitz, 1994; Schneider, 1994).

The distortion is locally specified by the two
parameters $\gamma, \varphi$ (or $\gamma_1, \gamma_2$).  To detect
this we follow the approach of Tyson, Valdes and Wenk (1990): we identify faint
galaxies and
form from the trace free parts of the quadrupole moments of their
images a two component entity that we call, following Blandford \etal, 1991,
the {\sl polarisation\/} $e_i$.  The relation between $e_\alpha$ and
galaxy shape is illustrated in figure \ref{fig:ellipses}.  The key
feature of this statistic is that in the absence of lensing it
averages to zero for statistically isotropic (i.e~randomly
oriented) objects, but in the
presence of lensing develops an expectation value that is
proportional to the gravitational shear.
There are many other ways one could detect this anisotropy, but
the quadrupole moment method seems to be practical and has so far
been used exclusively by all groups mentioned above (though with
subtle differences in how the technique is actually implemented;
see discussions of Kochanek, 1991, Miralda-Escude, 1991b,
and Bonnet and Mellier, 1994).

\begin{figure}
\myplotone{ellipses.ps}
\caption{Polarization values for a family of gaussian ellipsoid
objects of varying degrees of polarisation and orientation.
The equal area ellipses are contours of surface brightness.
The polarization values here are
$e_1 = (Q_{11} - Q_{22}) / (Q_{11} + Q_{22})$,
$e_2 = 2Q_{12} / (Q_{11} + Q_{22})$ where $Q_{ij}$ is a
simple unweighted central second moment.  For technical
reasons we actually use a weighted moment, but
at a qualitative level, the relation between shape and polarization
values is as shown here.}
\label{fig:ellipses}
\end{figure}

In real observations there will also be artificial distortion of the
images arising in the atmosphere and telescope.  These effects are of
two types: There may be a general distortion of the field arising in the
optics of the telescope and/or CCD camera.  This, like the gravitational
effect, consists of a stretching of the image. This can easily be measured from
the displacement
of stellar images and can be corrected for.  A more pernicious effect
is the {\sl smearing\/} of images with an anisotropic point spread
function (psf).
There are many possible sources of psf anisotropy; some of these
are discussed in appendix \ref{sec:sources}.
The signals one is trying to measure are very small; typically a few
percent in the outskirts of clusters, and around 1\% for
large-scale structure.  This kind of precision is possible in
principle due to the extremely large number of background galaxies
over the scale on which the shear is coherent: typically thousands for
clusters and potentially hundreds of thousands for large-scale structure.  It
is clearly
vital that systematic effects of this kind, which can easily
be comparable or greater than the signal, be corrected for.  As we shall see,
this appears to be quite feasible due to the presence of foreground stars,
which
provide a control sample from which one can measure the psf
quite precisely.

Over the past few years we have developed software which measures the
statistical anisotropy of faint galaxies and we have applied this
with some success to a number of clusters.
The purpose of this paper is to
describe the procedure in some detail, and in one place,
and to demonstrate that the techniques actually work
(though they may not yet be optimal).
The logical order of the paper follows the steps in the
analysis of actual CCD images.  In \S\ref{sec:detection} we describe the
object detection algorithm and in \S\ref{sec:objectanalysis}
we describe the galaxy photometry and shape analysis
algorithms. In \S\ref{sec:psfcorrection}
we describe how we can remove the effect of psf anisotropy
and finally in \S\ref{sec:shearmeasurement} we describe how
we quantify the shear.  All of the steps are illustrated with real
images, and in sections \ref{sec:psfcorrection} and
\ref{sec:shearmeasurement} we use deep HST images
which are smeared and stretched to simulate the effect of the
atmosphere, psf anisotropy and gravitational lensing.  These images
are then further degraded with noise to the same level as our typical
ground based data and
are then analysed in exactly the same manner.  These experiments
clearly demonstrate that we can measure and remove the effect of any psf
anisotropy
with high precision and they give a direct calibration of the effect
of seeing which previously required some modelling of the unknown
sizes and shapes of these faint and typical poorly resolved objects.

\section{Object Detection} \label{sec:detection}

Our method for detecting faint objects is very simple
and consists in essence of smoothing the
images and locating peaks. This is rather different
from e.g~the FOCAS approach (Jarvis and Tyson, 1981) which locates connected
regions which lie above a threshold, but is similar to some
other detection schemes (Kron 1980; Yee, 1991).

Our first attempts used a single fixed smoothing filter with
radius somewhat larger than the seeing disk; this is also the
strategy adopted by Bonnet and Mellier, 1994.  We experimented with
various shapes for the kernel, and opted for a simple gaussian
(Bonnet and Mellier use a `mexican-hat' filter).
Our algorithm has subsequently evolved, motivated in part
by considerations of a simple model in which one has gaussian
ellipsoid objects (plus noise) whose axial ratios and orientations one wishes
to measure by means of gaussian-window weighted quadrupole
moments.  It turns out that for a single isolated object of
this kind, the optimum scale for the filter is just equal to the
object's post-seeing scale length.
We also found that the signal/noise ratio was very sensitive to the choice
of smoothing radius, so it seems desirable that our
object detector should provide some reasonable estimate of the
object size to which we can tune our shape measurement window function.
Now for gaussian ellipsoids, the same filter radius
is also optimal for detecting the object (i.e.~the
signal/noise ratio for the peak is greatest when viewed at
this resolution), so if one smooths an image with all possible smoothing radii
and then chooses the peak of greatest significance,
this will provide both the position and the optimal filter radius
for the object. The significance here is defined to be
$\nu = {\rm S/N} =\propto f_s r_f$ where $f_s$ is the smoothed
peak surface brightness and $r_f$ is the filter radius.
Real galaxies are not gaussian of course, but the
sensitivity of signal to noise to the choice of weighting function is probably
a
general feature, and the hope is that the scale size
determined in this way can be used to improve the precision of our
shape estimation.

We have implemented such an algorithm on the computer.  We smooth the image
with
a range of filters, typically with steps in log filter radius of
 $\delta \ln r_f = 0.2$, find the peaks of the
smoothed images, and then link these together and construct a
catalogue of peak trajectories.  The behaviour of these
trajectories is illustrated in 1-dimension in figure \ref{fig:peaks1d}.
We initially used a simple gaussian smoothing kernel, but found that this
missed
some faint objects with bright neighbours, and we now use
a compensated `mexican-hat' style filter, though we have not explored
the possibilities here in great detail.

\begin{figure}
\myplotone{peaks1d.ps}
\caption{Peak trajectories for one dimensional model image consisting
a number of gaussian profile objects of varying sizes and central
surface brightnesses.  The model image is shown with and without noise
on the left, and the upper right panel shows the peak trajectories as
a function of smoothing radius (vertical scale).
Solid and dotted lines show maxima and minima and the
squares show the local maxima of the significance.
This shows the
generic behaviour of peaks in 1-dimension; at small smoothing radius
we have a large number of (mainly noise) peaks, but these gradually
annihilate with minima (where the smoothed image has an inflection
point).  In two-dimensions peaks annihilate with saddle points.
In the lower right panel are shown the local significance maxima
where the vertical scale is now the significance level
and the
width of the symbol denotes the smoothing radius.  The original
model image is shown superposed for comparison.  There is a clear
gap between the noise peaks and the real objects.  The algorithm
has detected all four real objects, and has assigned reasonable values
for their radii.  It has also detected the central pair of objects
as a significant object in its own right, and similarly
in two dimensions
we find that our `hierarchical-peak-finder' finds pairs and groups
of galaxies as well as individual galaxies.
Such composite objects can easily be removed from the final
catalogue if desired.}
\label{fig:peaks1d}
\end{figure}

\begin{figure}
\myplotfiddle{peaks2d.ps}
\caption{Hierarchical peak finding in 2-dimensions.
The panel on the left shows the raw output of the peak finder.
The image underlaid here is a small fragment
of the `avsigclip' sum of ten 20 minute
V-band exposures taken at the NTT with 0.7'' FWHM  seeing and with
0.34'' pixels.  The 256 by 512 pixel subimage shown here lies
a few arcminutes off the centre of A1689.
Most of
the obvious real objects are detected, but so are a large number
of (typically very small) noise peaks and a few apparent groups etc.  The panel
on the right
shows the result of applying a spatial coincidence test between
the raw catalogue shown on the left and a catalogue
constructed from a similarly deep I-band image of the same field.
Most of the small noise peaks have gone, as have most of
the composite objects. We also lose a few significant (in V)
but extremely blue objects.
The effective limiting magnitude in the final catalogue
is around I=25, and there are about 70 objects per square arcminute.}
\label{fig:peaks2d}
\end{figure}

The behaviour of the object finder in 2-dimensions
is shown in figure \ref{fig:peaks2d}.
The raw catalogue produced by the hierarchical peak finder was initially
filtered with a low significance threshold (3-sigma), and contained,
in addition to the real objects,
a large number of unwanted noise peaks
and some groups of galaxies.  We have found
an effective way to clean up the catalogue is to split the image data into
two subsets (usually we have some large number of images) and combine
these to form two images of the same patch of sky, but with
statistically independent noise properties.  Each combined image has a slightly
lower signal to noise
than would be obtained from the complete data set, but by setting a low
threshold we find we can still recover a high density of objects. We can then
remove cosmic rays and noise peaks etc.~by applying a spatial coincidence
test.  In the example shown in figure \ref{fig:peaks2d} we require
that the positions should coincide to within 2 pixels.  This effectively
removes the larger groups, as their positions tend to be more
uncertain than this, and also removes essentially all of the noise
peaks.
Once we have pruned our object catalogue in this way, we can of course
perform the photometric analysis of the objects using an image formed from
all the data (or simply average the properties of the
catalogues from the individual data sets).

We find the visual appearance of the final catalogue
in figure \ref{fig:peaks2d} quite encouraging. The
algorithm has detected essentially all of the objects of any significance, and
seems
to have assigned sensible scale sizes.
Our earlier attempts at object finding with a fixed smoothing filter
produced, in addition to most of the real objects, a large number
of false detections sitting on the diffuse
light around bright stars and foreground galaxies, and these had to be manually
edited from the catalogue.
In the near future we anticipate gathering
much larger images using large format CCD mosaics's (e.g.~Cuillandre \etal,
1994;
Luppino \etal, 1994), and having
a nearly automated data reduction pathway from image to
usable catalogue will obviously be a great practical advantage.
The algorithm is, aside from the choice of smoothing window shape, entirely
parameter free.
Finally, we mention that any detection algorithm like this
which uses a circularly symmetric smoothing kernel will be biased
in the shapes of the faintest objects as a circular object
will be more easily detected than an elongated one. As we shall see, this
does not seem to cause serious problems.

\section{Object Analysis} \label{sec:objectanalysis}

Armed with a catalogue of object positions we now determine a
number of shape, luminosity etc.~parameters.
Quite often we find that the preliminary data reduction leaves
some unwanted low-level low spatial frequency variations in the sky background.
This does not seriously affect the object detection phase, as we use a
compensated filter, but can have an effect on the shapes.  If we find
such problems we instruct the analysis software
to perform a local modelling of the sky background level.
This is illustrated in figure \ref{fig:sky}. Having corrected the
pixel values we are now ready to estimate basic photometric
parameters and shapes.

\begin{figure}
\myplotfiddle{sky.ps}
\caption{To determine the local sky level in the
vicinity of an object we first
determine the modal sky value in four sectors surrounding the object.
We then fit a bilinear model to these values.  The sectors are
illustrated for a subsample of the objects detected from the
NTT data. This step is optional.}
\label{fig:sky}
\end{figure}

\subsection{Basic Photometry} \label{sec:basic photometry}

The object finder already provides a crude estimate of the radius and
luminosity of the objects.
To determine a better half-light radius, which we use for star/galaxy
separation,
and `total' magnitude we proceed as
follows.  Using the surface brightness corrected for the local background
we calculate the growth curve for the integrated light
$l(<r) \equiv 2\pi\int d\theta\;\theta f(\vectheta)$ (where $f(\vectheta)$
denotes the surface brightness)
as a function of radius and determine a half-light radius and
total luminosity within an aperture.  We have experimented
with different types of apertures.  We currently favour an aperture
of three times the scale length returned by the object finder; these
are the circles drawn in figure \ref{fig:peaks2d}.  We have also
used an alternative which is three times
the Petrosian radius (this is the radius where $l(<r)/r$ peaks).
The factor three in both cases being chosen as subjective
compromise between obtaining a `true' total magnitude and precision.
The aperture radii are generally very similar, but for some
of the objects the Petrosian radius estimator appears unreliable.
Examples of the results of the basic photometric analysis are shown
in figure \ref{fig:a1689}.

\begin{figure}
\myplotone{a1689.ps}
\caption{Upper panels show
half-light radii and magnitudes determined from
typical ground based observations.  The data here were taken at the
NTT, with 0.7'' seeing and 0.34'' pixels.  The vertical stellar locus is
clearly
seen, and it is possible to separate moderately bright stars from galaxies with
confidence --- this is vital for measuring and correcting for point spread
function anisotropy.  The bright stars are saturated and consequently
swell up, but the transition seems quite sharp and it appears to be quite
easy to select a sample of moderately bright stars with which one
can reliably measure the psf.
The lower panel shows the color magnitude diagram for extended
objects; the cluster (A1689)
sequence is very sharply delineated.}
\label{fig:a1689}
\end{figure}

\subsection{Shape Estimation} \label{sec:shapeestimation}

The shape parameters we use are formed from weighted quadrupole moments
\begin{equation}	\label{eq:Qijdefinition}
	Q_{ij} \equiv \int d^2 \theta W(\theta) \theta_i \theta_j f(\vectheta)
\end{equation}
where angles are measured relative to the object position as determined
in the detection phase.
We take $W(\theta)$ to be a gaussian with scale length equal to some
multiple of the scale determined in the object detection phase.  Currently
we use a multiplier of unity, as this seems from experimentation to
be a good choice --- with more extensive HST data and experiments of the
kind performed below it should be possible to optimise this
parameter.

We then define the polarisation parameters by
\begin{equation}	\label{eq:edefinition}
	e_\alpha \equiv Q_\alpha / T
\end{equation}
with
\begin{equation}	\label{eq:Q1Q2Dedefinition}
\begin{array}{c}
Q_1 \equiv Q_{11} - Q_{22}\\
Q_2 \equiv 2 Q_{21}\\
T \equiv Q_{11} + Q_{22}
\end{array}
\end{equation}
which clearly provides some measure of the ellipticity of an object:
$e_\alpha = 0$ for a circularly symmetric object.  Now under a rotation
of the coordinate frame $e_\alpha \rightarrow
R_{\alpha\beta}(2\varphi) e_\beta$, where $R$ is the 2-dimensional
rotation matrix, so, in
the absence of lensing, the polarisation values will be isotropically
distributed about the origin on the $e_1,e_2$ plane.
In the case of unweighted moments ($W=1$) it is
easy to see how these statistics will be perturbed by a gravitational shear,
since the mapping
$f(\theta) \rightarrow f(\psi \cdot \theta)$ simply corresponds, in the
diagonal frame, to a rescaling of the coordinate axes so
$Q'_{ij} = Q_{ij} / \Lambda_1 \Lambda_2 \Lambda_i \Lambda_j$,
or, in the weak shear limit,
$Q'_{ij} = Q_{ij}
( 1 - (\lambda_1+ \lambda_2 + \lambda_i + \lambda_j))$, where as before primed
and unprimed quantities denote the perturbed
and unperturbed values respectively.
For isotropic objects $\langle Q_1 \rangle = 0$, but in the presence of shear
there will be a systematic shift in $\langle Q'_1 \rangle$
proportional to $\lambda_1 - \lambda_2 = \gamma$.
The way we have chosen to normalise the moments (to unit trace), the shift
in the polarization due to a given shear depends to a slight extent on the
polarization of the object (and so the shift in the mean polarization
depends somewhat on the distribution of ellipticities).
Bonnet and Mellier (1994) have made the interesting
suggestion that one normalise to unit determinant.  The shift in the
polarization
is then independent of the intrinsic ellipticity; in fact the mean shift
is just $\langle \delta e_\alpha \rangle = 2 \gamma_\alpha$.
This simple relation between $\delta e$ and $\gamma$ is a nice
property, but unfortunately
this does not hold for weighted moments (and unweighted second
moments are impractical due to divergent noise).
With weighted moments and with either normalisation scheme, the
shift depends in a non-trivial way on the intrinsic shapes distribution,
but, as we will see,  this does not present an insurmountable problem.

If one ignores, for the moment, the effects of seeing and photon
counting noise,  it is fairly straightforward
to calculate how the polarization parameters will change under a
constant (and small) gravitational shear for an arbitrary window
function $W(\theta)$.
As we show below, the 1st order change
in polarization induced by the shear can be written as
\begin{equation}
\delta e_\alpha = P^\gamma_{\alpha\beta} \gamma_\beta
\end{equation}
where $P^\gamma_{\alpha\beta}$  defines the {\sl shear polarizability
tensor\/}.
Note that,
strictly speaking, $P^\gamma_{\alpha\beta}$
is not a tensor: $e_\alpha$ and $\gamma_\alpha$ are two
component entities, but they are not vectors as they
transform under rotations as
$e_\alpha \rightarrow R_{\alpha\beta}(2 \theta) e_\beta$. Similarly,
our polarizabilities transform as
$P_{\alpha\beta} \rightarrow R_{\alpha\gamma}(2 \theta) P_{\gamma\delta}
R_{\delta\beta}^{-1}(2 \theta)$.  We will henceforth sloppily refer to
these objects as tensors and vectors, but we try to distinguish them by
using greek symbols for their
indices and using latin indices for real vectors and tensors.

Now $P^\gamma_{\alpha\beta}$ is some rather messy combination of
angular moments of the surface brightness,
but the important point is that it can be directly measured
for each individual galaxy image, and so provides a way to calibrate the
polarization
statistics: the average of $e_\beta / P^\gamma_{\alpha\beta}$ over the galaxies
lying on some patch of sky is
just proportional to the shear $\gamma_\alpha$ averaged over
the same region. In this way we can always construct a statistic
which gives an unbiased estimate of the shear.
Now noise in the image will mean that
we make some error in calculating $P^\gamma_{\alpha\beta}$, but, as we
will see, this does not seem to be a serious problem.  A bigger problem
comes from seeing, which will perturb both the polarization and the
polarizability in a systematic way.  The approach we have adopted
is a semi-empirical one:  We first calculate as best we can the polarizability
ignoring seeing --- this should provide a set of shear estimates which have a
negative bias which depends on the image size, but which should asymptote
to the correct value for large images --- and we then empirically
calibrate the  seeing induced suppression as a function of image size
by using HST data which we artificially stretch and then degrade to simulate
ground based observing conditions.
The technical details of this are given below (\ref{sec:shearmeasurement}).

First however, we will address a closely related problem: how the mean
polarization
shifts in response to smearing with an anisotropic psf and how we can,
by measuring the psf of foreground stars, annul this.  Provided the
psf is close to circular --- and this fortunately seems to be the
case for the data we have looked at --- we can define
a linearised `smear polarizability' such that
\begin{equation}
\delta e_\alpha = P^s_{\alpha\beta} p_\beta
\end{equation}
where $p_\beta$ is some measure of the psf anisotropy.
The smear polarizability can, like $P^\gamma_{\alpha\beta}$, be calculated
for each object. The nice thing here is that $P^s_{\alpha\beta}$
depends only on the image shape {\sl after\/} seeing, and so can be
calculated exactly, in the absence of noise, and seems to be a rather
robust statistic even in the presence of noise for our faint galaxies.
The bulk of the rest of the paper is devoted to the calculation of these
polarizability tensors, and to a description of experiments
with the HST observations to show how they work under realistic
conditions with noise, finite pixel size,
crowding of neighbouring images etc.

\section{Correction for psf Anisotropy} \label{sec:psfcorrection}

A number of sources of psf anisotropy are discussed briefly in appendix
\ref{sec:sources}. Most of these produce an anisotropy which
is constant across a CCD frame
or varies in a smooth manner. Provided one has sufficiently many foreground
stars --- and they must be not too bright that their shapes are distorted by
non-linearity in the CCD or readout electronics --- one can map
the psf anisotropy.
One way to correct would be to reconvolve the image with an artificial
psf designed to give a circular final psf, but this involves
some loss of information.  The approach developed here is to calculate how the
polarization values of the galaxies respond to
a given psf anisotropy and apply an appropriate correction. We first
present the analysis and then demonstrate the procedure with
realistic test data.

\subsection{Analysis} \label{sec:psfanalysis}

As we show in appendix \ref{sec:smear},
one can model any source of psf anisotropy as a convolution of a
circularly smeared image with a small, but highly anisotropic
kernel $g(\vectheta)$ (in many cases this is just a small uni-directional
smearing: $g(x,y) = \delta(x) g'(y)$ with some box-car like function $g'$).
For small psf anisotropy, the shift in the polarization $e_\alpha$
depends only on $p_\alpha \equiv \{q_{11} - q_{22}, 2 q_{12} \}$  where
\begin{equation}
q_{lm} \equiv \int d^2 \theta \theta_l \theta_m g(\vec \theta)
\end{equation}
is the unweighted quadrupole moment of $g$, and we have assumed that
$g$ is normalised such that $\int d^2 \theta g = 1$ and that
the origin of coordinates is chosen so that $\int d^2 \theta \theta_i g(\vec
\theta) = 0$.
The perturbation to the polarisation is, to linear order
in $p_\alpha$,
\begin{equation}
\delta e_\alpha = P^s_{\alpha\beta} p_\beta
\end{equation}
where the
smear polarizability tensor is
\begin{equation}	\label{eq:psmear}
P^s_{\alpha\beta} = X^s_{\alpha\beta} - e_\alpha e^s_\beta
\end{equation}
where
\begin{equation}	\label{eq:Xsmear}
X^s_{\alpha\beta} = {1\over T}\int d^2 \theta
\left[
\begin{array}{cc}
2 W + 4 W' \theta^2 + 2 W'' (\theta_1^2 - \theta_2^2)^2  &
4 W'' (\theta_1^2 - \theta_2^2) \theta_1 \theta_2 \\
4 W'' (\theta_1^2 - \theta_2^2) \theta_1 \theta_2 &
2 W + 4 W' \theta^2 + 8 W'' \theta_1^2 \theta_2^2
\end{array}
\right] f(\vectheta)
\end{equation}
and
\begin{equation}	\label{eq:esmear}
e^s_\alpha \equiv {1\over T}\int d^2 \theta
\left[
\begin{array}{c}
\theta_1^2 - \theta_2^2 \\
2 \theta_1 \theta_2
\end{array}
\right] (6 W' + 2 W'' \theta^2) f(\vectheta)
\end{equation}
and where prime denotes differentiation wrt $\theta^2$.
The surface brighness here is that after any circularly
symmetric seeing, which to zeroth order in $p_\alpha$ is
just the observed surface brightness.
Note that for the impractical case of an unweighted quadrupole moment
($W' = W'' = 0$) the smear polarizability is diagonal with
$P^s_{11} = P^s_{22} = \int d^2 \theta f / \int d^2 \theta \theta^2 f
= \langle \theta^2 \rangle^{-1}$
independent of the intrinsic ellipticity distribution. This would
not be the case were we to normalise to unit determinant.
Note also that $P^s_{\alpha\beta}$ is diagonal for any circular object
such as a star.

We can measure
the shear polarizability for each individual object.
The stars then provide an estimate of
$p_\alpha = e_\alpha /
P^s_{\alpha\alpha}$ (no summation),
and we can correct each galaxy polarization by an amount $-P^s_{\alpha\beta}
p_\beta$
and restore the polarization values to what would have been seen with a
perfectly isotropic psf.
All this assumes noise-free data. The stars are reasonably bright, so
photon counting noise is little problem there. The polarizabilities estimated
for the galaxies will be much noisier, and this introduces both
a random and systematic error in $P^s_{\alpha\beta}$.  The random error is
relatively
benign, since our goal is to finally determine a shift in the
mean polarisation.  The systematic error arises in the second term in
$P^s_{\alpha\beta}$ which is quadratic in the surface brighness, so for very
faint
objects this will introduce a bias as the observed $ e_\alpha
e^s_\alpha$ will be inflated by photon counting noise.
One could easily calculate the size of this effect and apply
an appropriate correction, but in fact
the second term tends to be quite small, and we have not done so.

For the correction procedure to work it is vital that one has
sufficiently many stars to sample any variation of the psf anisotropy
across the chip. It the case of the NTT data shown in figure \ref{fig:peaks2d}
we obtained a subsample of about 30 stars (on $\simeq$ 70 square arcmin)
which were sufficiently bright that they can be distinguished
from galaxies with near certainty yet not so bright that they are
saturated. As discussed in appendix \ref{sec:sources}, most
anticipated effects will produce a slowly varying
anisotropy, so the number of stars should be adequate.
In a typical observing run we end up with some number of
fields (each comprising the `avsigclip' sum of several images).  From the
subsample of stellar objects
our software currently attempts to fit a model in which there is a
spatially constant anisotropy $\overline p_\alpha$ for each
field (this should accurately describe influences such as wind shake
and atmospheric dispersion) and a `global' low order polynomial in
angle on the chip which should accommodate any reproducible
aberration effects.  The software starts with a zeroth order
fit and then increments the order of the fit while monitoring the
residuals.  As an example, for the Fahlman \etal\ (1994) data we found a
significant linear gradient but no significant improvement in fit
was obtained for higher order fits, so we used the linear model.

\subsection{An Experiment with HST Data}	\label{sec:psfexperiment}

The expression (\ref{eq:psmear})
for the smear polarizability is rather involved,
to say the least, so it is difficult to analytically quantify
the uncertainty and the various biases present in the psf correction
process.  One way to rigorously test the procedure is to
take very deep images and convolve
them with a small but highly anisotropic psf so the total
psf develops a small anisotropy.  We can then add noise
to the level in our typical integrations and analyse
the images, separate and measure the polarizations of the stars, and then
apply the appropriate correction.
We can then see how well the correction works in a realistic situation.
It is necessary that we start
with very long integrations however, since we would otherwise be
smearing any noise in the original image, which would be unrealistic.

It turns out that HST data are very useful here. While the aperture
of the telescope is
relatively small, the sky is so much fainter for HST (about 2.5 magnitudes
in the I band) that one can go very deep in integrations of reasonable
length. The data we will use here are $\sim 2$ hrs WFPC2 integration on a
single
target.  This gives us 3 CCD frames, each $1.25'$ square.  Part of
one of these is shown in figure \ref{fig:HST}.
The rms level of the noise added was about twice that in
the rebinned (but unsmoothed) HST data, so the noise from
the original image should be negligible.  The limiting magnitude
in the degraded images is very similar to that in the NNT data
shown above.  The total area is 5 square arcmin yielding
a few hundred detectable galaxies.

A further advantage of using
HST data is that it allows one to address the question of ``pixelization''.
The expressions for the smear polarizability etc.~are all expressed
as integrals of continuous functions, whereas in reality they are
implemented as discrete sums over pixels.  The $0.1''$ WFPC2 pixels
are much smaller than the pixels in the rebinned images, so any
effect arising from the finite pixel size in the ground based
observations should be seen in the simulations.

\begin{figure}
\myplotfiddle{HST.ps}
\caption{On the upper left is shown roughly a quarter of a single
WFPC2 field (380 by 380 pixels at 0.1'' per pixel), and below it the result
of smoothing this to simulate 0.5'' seeing and rebinning
to 0.2'' pixels. This is a 3 orbit I-band integration. In the
upper right panel we have added noise
to the level appropriate for a similar length integration on a telescope like
the
CFHT or the NTT. For each field,
two such degraded images (with independent realisations of noise)
were analysed in exactly the same way that we analyse the real
ground based data, and the chart on the lower right shows the result of the
object finding after spatial coincidence testing.  The final density
of objects is very similar to those found in the NTT data shown above.}
\label{fig:HST}
\end{figure}

Now to test the psf anisotropy correction machinery we have degraded
these data much as shown in figure \ref{fig:HST}, but with an anisotropic
psf: a gaussian ellipsoid with $a/b \simeq 2$ and with the same area as
the psf used in figure \ref{fig:HST}.  In an attempt to
to boost the statistical signal somewhat we have
smeared the images in four ways with psf position angles 0,45,90,135 degrees,
and we have used two independent realisations of noise, though the results
are not really statistically independent.
The psf induced polarisation is clearly seen in the upper left
panel of figure \ref{fig:smearbefore} (for each frame we have applied
the rotation matrix $R_{\alpha\beta}(2\varphi)$ to the $e_\alpha$ estimates
so that the $\theta_1$
direction lies along the stellar major axis).  There is considerable
`cosmic variance' in this plot due to the random intrinsic
ellipticities of the relatively small number of real galaxies used.
We can get a somewhat
cleaner picture of the shift if we pair up objects found from
the anisotropically smeared images with those found with
circular seeing, and plot
the {\sl change\/} in the polarisation
values $\delta e_\alpha$ introduced by the psf anisotropy.
This is shown in the upper right panel.  The lower panels in figure
\ref{fig:smearbefore} show that the induced ellipticity
shift is primarily a function of the image size,
being greatest for the smallest images as expected, and has a rather weak
dependence on luminosity.  Figure \ref{fig:smearafter} shows the same plots,
but after measuring the shear from the actual stellar images and applying
the correction as described. Clearly the procedure has worked very well
indeed, and any residual anisotropy is very small.

There is a hint that
the method may overcorrect the faintest galaxies.  The catalogues used
here were limited to 5-sigma detections and above.  If one includes even lower
significance detections then the overcorrection appears to be stronger.
Perhaps this is because we are seeing the bias towards circular objects at
very low significance level.  There might also be some residual arising from
departures from linearity as the psf was so strongly anisotropic, but
the limited numbers of galaxies here do not allow us to measure
this with any precision.

\begin{figure}
\myplotone{smearbefore.ps}
\caption{Upper left panel shows the polarization values
for images smeared with an anisotropic psf.  Upper right
shows the changes in polarization for smeared/unsmeared pairs of images.
Lower panels show the dependence of the polarization shift on
radius and apparent magnitude. The solid line is a simple
moving average.  The expected inverse trend with
image size is clearly seen.}
\label{fig:smearbefore}
\end{figure}

\begin{figure}
\myplotone{smearafter.ps}
\caption{Polarization and polarization shift distributions after
correction for psf anisotropy as described in the text. The degraded
HST data were analysed in exactly the same way we analyse the
real ground based data, so this provides a rigorous and exacting
test of the method which it seems to have passed very well.
If we average over a 2-magnitude range as we typically do with
the real data then any residuals are less than about 10\% of the
uncorrected induced polarisation.
There is a hint that the method may overcorrect the faintest
objects, but more data are needed to establish if this is a real
effect.}
\label{fig:smearafter}
\end{figure}

The issue of psf anisotropy correction has been considered by Bonnet and
Mellier
(1994), and by Mould \etal, 1994, but both groups have derived
a correction different from ours.
Bonnet and Mellier treat the effect of psf anisotropy
as though it were a stretching of the images rather than a smearing.   The
correction
that they apply is then independent of the image size.
It is intuitively reasonable that a smearing (unlike a stretching) will produce
an image polarization which scales in inverse proportion
to the {\sl area\/} of the image
and this is revealed quantitatively in our analysis:
one can see on dimensional grounds that the smear polarizability
defined by equations \ref{eq:psmear},\ref{eq:Xsmear},\ref{eq:esmear}
is a measure of the inverse area of the image, and the strong
dependence of psf induced polarisation on image size is shown graphically in
figure \ref{fig:smearbefore}.
As the Bonnet and Mellier correction is designed to cancel any anisotropy in
the stars,
we would expect that the result will be to  overcorrect the galaxy
polarisations.
We would certainly make a serious error were we to apply a constant correction,
but the kernel that Bonnet and Mellier use is rather different from ours, and
so one would have to redo the analysis described above in order to quantify the
error. Mould \etal 1994 argue that the effect of a psf anisotropy will be a
polarisation which is inversely proportional to the linear size of the
images rather than the area as we find here.  Now they were using the FOCAS
software which
measures the second moments within an isophotal aperture which may behave
differently from the weighted moments we use, but we suspect their procedure
also over-corrects the galaxies.

It is interesting to ask: which is the dominant source of noise in the
mean polarization, the
$\sim e_{\rm rms} / \sqrt{N}$ fluctuations arising from the
random intrinsic galaxy ellipticities and measurement error
averaged over a large number of background galaxies, or the
error feeding through from any error on the psf anisotropy, which
is determined from a small number of stars?
It is difficult to give a definitive answer
to this question, as it depends on how large is the systematic
psf anisotropy one is trying to correct, but our experience is that the
dominant source of noise in the final error budget
comes from the intrinsic variance in galaxy shapes.
Note that if one is taking large numbers of images to get high
signal/noise on a particular target lens then any stochastic
psf anisotropy will tend to average away, and systematic effects
should be removable in principle by e.g.~rotating the CCD by $\pi/2$ between
pairs of images taken under otherwise identical conditions, and
the noise from the random intrinsic ellipticities must, under
such conditions, eventually come to dominate.

\section{Measuring the Shear} \label{sec:shearmeasurement}

Having removed the effect of psf anisotropy we are now ready to
extract a quantitative estimate of the shear.
As we show in appendix \ref{sec:shear}, if the galaxies are well resolved
it is relatively straightforward to calculate how the polarization
values are perturbed by a gravitational shear.  Much as in
\S\ref{sec:psfcorrection} we obtain a linear response
\begin{equation}
\delta e_\alpha = P^\gamma_{\alpha\beta} \gamma_\beta
\end{equation}
but now with
\begin{equation}
P^\gamma_{\alpha\beta} = X^\gamma_{\alpha\beta} - e_\alpha e^\gamma_\beta
\end{equation}
with
\begin{equation}
X^\gamma_{\alpha\beta}= {1\over T}\int d^2 \theta
\left[
\begin{array}{cc}
2 W \theta^2 + 2 W'(\theta_1^2 - \theta_2^2)^2  &
4 W' (\theta_1^2 - \theta_2^2) \theta_1 \theta_2 \\
4 W' (\theta_1^2 - \theta_2^2) \theta_1 \theta_2 &
2 W \theta^2 + 8 W' \theta_1^2 \theta_2^2
\end{array}
\right] f(\vectheta)
\end{equation}
and with
\begin{equation}
e^\gamma_\alpha = 4 e_\alpha + {2 \over T} \int d^2 \theta
\left[
\begin{array}{c}
\theta_1^2 - \theta_2^2 \\
2 \theta_1 \theta_2
\end{array}
\right] \theta^2 W' f(\vectheta)
\end{equation}
where prime denotes differentiation wrt $\theta^2$.  For the impractical
but simple case of constant weight we find
$P^\gamma_{\alpha\beta} = 2(\delta_{\alpha\beta} - e_\alpha e_\beta)$ and
$\langle P^\gamma_{\alpha\beta} \rangle = 2 \delta_{\alpha\beta}
(1 - \langle e_\alpha e_\alpha \rangle)$.

A fair estimate of the shear $\gamma_\alpha$ is therefore given by
taking the mean of $e_\alpha P^{\gamma -1}_{\alpha\beta}$;
the shear polarizability
effectively providing a shape dependent calibration factor.
Unfortunately all this ignores seeing which will dilute the
polarization and also modify the polarizability tensor,
and calculating the appropriate correction requires
knowledge of the pre-seeing galaxy shape.  The approach we
have adopted in the past is as follows: We first calculate the shear
polarizability
and estimate $\gamma_\alpha$ as above.  This will underestimate
the true shear, but should at least asymptote to the correct
value for large galaxies.  We then empirically determine a correction
as a function of galaxy size. In Fahlman \etal, 1994 we estimated the
correction under the assumption that the faint galaxies are
scaled down replicas of their brighter (and better
resolved) cousins.  Here we make a more direct estimation
by taking HST data, shearing them and
then degrading them to simulate terrestrial observing conditions.

An alternative would be to simply develop an empirical calibration
of the shear/polarisation relation without trying to calculate
the shear polarizability.  This might actually improve the precision
of the shear estimate because there appears to be a rather narrow
distribution of shear polarizabilities (this is not surprising because
$P^\gamma$
depends only on the shape of the galaxies, unlike $P^s$
which is an inverse measure of the area of the image and so
varies considerably from object to object), and the $P^\gamma$
values are rather noisy. Thus, when we divide by
the shear polarizability we may make a relatively large error in what
should really be a small correction.  From the experiments here it appears
that the precision is very similar for these two methods, so more
extensive experiments are required to see which is optimal.

\subsection{Empirical Calibration of the Effect of Seeing}

To empirically calibrate the shear estimator with finite seeing
we used
the same WFPC2 data as in section \ref{sec:psfexperiment}.
These images were stretched to simulate the effect of lensing
as shown in figure \ref{fig:shearedimages}.  We then
smeared these with a circular gaussian psf to simulate
seeing, rebinned the images to a pixel scale appropriate for
ground based observations, and added multiple realisations of noise.
We then applied the object finding algorithm
with coincidence testing as described in \S\ref{sec:detection}.
This experiment is arguably somewhat unrealistic in that we apply a
constant stretch to all of the objects whereas in reality we
have a superposition of source `screens' each being smeared
by a different amount. As the goal here is simply to determine
a calibration factor for the shift in the mean polarization
this should not be much of a problem, and in any case,
for cluster lenses at the redshifts
we currently target ($z \lesssim 0.2$), and for reasonable
magnitude limits to define the faint galaxy subsample, the constant shear
approximation
is quite good as we show in figure \ref{fig:betadistribution}.

\begin{figure}
\myplotone{beta.ps}
\caption{Distribution of distortion strengths for a realistic distribution
of background galaxy redshifts and for various lens redshifts.
The lower panel shows (solid) the smoothed redshift distribution for the
I=20-22 CFRS redshift survey (Lilly, \etal, 1994) which is nearly
complete and represents an increase of about an order of magnitude
over previous surveys at these magnitudes (Lilly, 93, and Tresse \etal\ 1993).
The dashed curve is an extrapolation to fainter magnitudes I=22-24 made
assuming
no evolution and kindly provided by Simon Lilly.  The upper panels show, for
these two magnitude slices,
the distribution of  \protect{$\beta$} values --- the distortion strength
relative to
that for an infinitely distant
object --- for 5 lens redshifts $z_l = 0.1,0.15,0.2,0.3, 0.5$
progressing from right to left.
The curves move progressively to the right and become more sharply
peaked as the lens redshift decreases.
For $z_l \lesssim 0.2$ and for the faint magnitudes used here the
$\beta$ distribution is very narrow and the single
source plane approximation should be acceptable.}
\label{fig:betadistribution}
\end{figure}

\begin{figure}
\myplotfiddle{stretch.ps}
\caption{Artificially sheared HST images. The original image here was
a single WFPC2 CCD field, and a
shear of 15\% was applied
to the CCD images in four directions as shown.  In total 3-frames
like this were used.}
\label{fig:shearedimages}
\end{figure}

As a test of the shear-polarizability calculation we then
analysed the objects found under realistic
conditions, but using the unsmeared (but stretched) images.
The result is shown in figure \ref{fig:noseeing}. The
image anisotropy is clearly seen, and calibration seems to
have recovered the correct amplitude as expected.
We now repeat the shapes calculation but using the stretched, smeared
and noise-added images; figures \ref{fig:seeing05}, \ref{fig:seeing07}.  The
expected dilution of the signal (primarily a function of
image size) is clearly seen.  With more extensive data it should
be possible to calibrate this effect and quantify the
trend with image size quite accurately.  The
experiment should give at least a reasonably
secure overall calibration factor for any particular
observation.

\begin{figure}
\myplotone{seeingoff.ps}
\caption{Test of  shear polarizability calculation.  The dashed
line shows the shear applied and the points show the individual estimates.
The solid line shows a simple moving average of the shear estimates.
The sample of objects analysed here were those detected from the
realistically degraded images, but the polarization was actually
measured using the unsmeared images.  This experiment verifies that
the analytic shear polarizability does indeed correctly
calibrate the raw polarization estimates.}
\label{fig:noseeing}
\end{figure}

\begin{figure}
\myplotone{seeing05.ps}
\caption{Shear estimates with FWHM = $0.5''$ seeing. The shear is still
detected at a high level of significance but the suppression --- primarily
a function of image size --- due
to seeing can be seen.  The radii here are in units of the $0.2''$
pixels of the rebinned images.}
\label{fig:seeing05}
\end{figure}

\begin{figure}
\myplotone{seeing07.ps}
\caption{As before, but with $0.7''$ seeing.  The dilution is now
stronger.}
\label{fig:seeing07}
\end{figure}

As well as showing the trend of mean polarisation with image
size, figures \ref{fig:seeing05},\ref{fig:seeing07}  also shows empirically how
the scatter about the mean trend increases for the fainter and
smaller objects.  This should allow us to devise an optimal
weighting scheme for combining the shear estimates, but this requires
more extensive HST data than is currently available.  Similarly,
one might consider a more sophisticated weighting scheme than simply averaging
the shear estimates.  We have experimented with this and with other
schemes such as trying to determine the mode of the polarization,
but none of these has yielded significant improvement over
a straight average (this is because the distribution of random
polarizations is quite close to a 2-dimensional gaussian and for
a gaussian the optimal weight is uniform).

\section{Summary}

Images of faint galaxies are subject to two weak
influences which cause their shapes to be polarized:  Tidal shearing of the
rays as they propagate through intervening clusters and
large scale structure, and smearing with an anisotropic
instrumental psf when they reach the Earth.
We have described software which allows
one to measure and correct for the latter to a high degree of precision,
and gives a quantitative estimate of the former, providing
a unique and direct probe of the total mass distribution
in the universe.

We have subjected the software to rigorous
testing using HST data which are stretched to simulate
lensing and then degraded to ground based resolution and
noise levels, and with an anisotropic psf.  Analysing
these data exactly as we do the ground based data we have
shown that the correction procedure developed here
appears to work very well: any residual polarization arising from
psf anisotropy is too small to detect with the small
number of galaxies here.  We emphasise that the correction
procedure derived here is different from the techniques developed
by Bonnet and Mellier (1994) and by Mould \etal\ (1994), but we
believe we have demonstrated the validity of our method
both analytically and in tests under realistic conditions.

The second main result of the paper is a demonstration of the calibration of
the
effect of seeing on our shear estimator. The HST data
play a vital role here, and we have shown how the suppression depends
on image size and brightness for typical observing
conditions (i.e.~pixel and seeing disk scale).
This removes a major question mark over the calibration of these
observations.
The limited data we use here are, we believe, sufficient to give
a reasonable mean calibration for a given choice of magnitude
limits.  With more HST data we can hope to develop a detailed model
for the dependence of the polarizability on size, luminosity etc.

The work described here could usefully be extended in several respects.
The polarizability tensor analysis we have developed is specific to the
choice of polarization estimate (normalised to unit trace) that we have
somewhat arbitrarily adopted.  It would be fairly
straightforward to perform the analogous calculations with
polarisation estimators normalised to unit determinant as
suggested by Bonnet and Mellier, 1994, but we have not
done so.
While our analysis applies for an arbitrary weighting function
the numerical results are specific to the case of a gaussian weight,
and would require recalculating for any other weight function.
There are a number of other parameters or features
of the scheme which have been set with little serious attempt
at optimisation.  The analytic machinery and experimental method
developed here provides a quantitative way to measure the performance
of any particular scheme.  However, to determine the
optimal point in this large parameter space will require more
extensive data than are currently available.
Regarding psf anisotropy, we are confident that our
procedure is adequate to remove any artificial anisotropy
at the level required for mapping the mass in clusters, but
are less certain about the more ambitious goal of measuring
large-scale structure, where the accuracy of
the correction is more critical.
Any {\sl achromatic\/} psf anisotropy
can be removed, and we suspect that any residual spectrum
dependent effects are
actually quite small, but a definitive calculation --- which would
depend on the details of the correcting optics in the telescope,
the choice of filter, the statistical distribution of
the spectra of stars and faint galaxies, and so on --- remains to be done.

\acknowledgements{We thank Derrick Salmon, Chris Pritchet, Keith Taylor,
Simon Lilly, Greg Fahlman for many enlightening discussions.}

\appendix






\section{Smear Polarizability} 		\label{sec:smear}

In this section we show how the polarization of an image $e_\alpha$
is perturbed by anisotropy of the psf. Using a linearised
analysis (valid if the psf is nearly circular) we derive a
`polarizability' $P^s_{\alpha\beta}$ such that $\delta e_\alpha
= P^s_{\alpha\beta} p_\beta$ where $p_\beta$ is a measure of the
psf anisotropy.

We discuss several sources of psf anisotropy in
\S\ref{sec:sources}.
We can model the effect of all of these as a
convolution of the post-seeing circularly smeared image with
a small, normalised, but highly anisotropic psf:
\begin{equation}
f'(\vec \theta) = \int d^2 \theta' g(\vec \theta')
f(\vec\theta - \vec\theta')
\end{equation}
This is obviously true for wind-shake and other effects
arising in the telescope, but is also valid for e.g.~atmospheric
dispersion, which happens at the same time as the circular
smearing, due to fact that convolutions commute.

We can set the spatial origin so that the 1st moment of $g$ vanishes, and we
find, on Taylor
expanding $f$, the
weighted quadrupole is perturbed according to
\begin{equation}	\label{eq:deltaQ}
Q'_{ij} \equiv \int d^2 \theta W(\theta) \theta_i \theta_j
f'(\vec \theta) = Q_{ij} + \delta Q_{ij}  = Q_{ij} + q_{lm} {\cal Z}_{lmij}
\end{equation}
where
\begin{equation}
q_{lm} \equiv \int d^2 \theta \theta_l \theta_m g(\vec \theta)
\end{equation}
and where
\begin{equation}
{\cal Z}_{lmij} = \int d^2 \theta W(\theta) \theta_i \theta_j
f_{,lm}(\vec \theta) = \int d^2 \theta f(\vec\theta) z_{lmij}
\end{equation}
where we have integrated by parts twice and we have defined
\begin{equation}
z_{lmij} = {\partial (W(\theta) \theta_i \theta_j) \over
\partial \theta_l \partial \theta_m}
\end{equation}

The simplest case is $W = 1$ in which case we have
\begin{equation}
z_{lmij} = \delta_{il}\delta_{jm} + \delta_{jl}\delta_{im}
\end{equation}
In the general case
\begin{equation}
z_{lmij} = W(\delta_{il}\delta_{jm}+ \delta_{jl}\delta_{im})
+ 2 W'(
\delta_{im} \theta_j \theta_l +
\delta_{jm} \theta_i \theta_l +
\delta_{il} \theta_j \theta_m +
\delta_{jl} \theta_i \theta_m +
\delta_{lm} \theta_i \theta_j
)
+ 4 W'' \theta_i \theta_j \theta_l \theta_m
\end{equation}
where prime denotes differentiation wrt $\theta^2$.

The trace of $q_{ij}$ will change the size of the image, but not
the shape, so we take $q_{ij}$ to be trace free and, of course, symmetric:
\begin{equation}	\label{eq:psfanisotropydef}
\left[
\begin{array}{cc}
q_{11} & q_{12} \\
q_{21} & q_{22}
\end{array}
\right]
= {1\over 2}
\left[
\begin{array}{cc}
p_1 & p_2 \\
p_2 & -p_1
\end{array}
\right]
\end{equation}
which defines the psf anisotropy vector $p_\alpha$.

{}From the definition of the polarization (equation \ref{eq:edefinition})
in terms of $Q_{ij}$ and using equation \ref{eq:deltaQ}
we have, to linear order in $p_\alpha$
\begin{equation} \label{eq:Psmear}
\delta e_\alpha = P^s_{\alpha\beta} p_\beta
\end{equation}
with
\begin{equation}
P^s_{\alpha\beta} = X^s_{\alpha\beta} - e_\alpha e^s_\beta
\end{equation}
where
\begin{equation}	\label{eq:Xsdef}
X^s_{\alpha\beta} = {1\over T} \int d^2 \theta
\left[
\begin{array}{cc}
2 W + 4 W' \theta^2 + 2 W'' (\theta_1^2 - \theta_2^2)^2  &
4 W'' (\theta_1^2 - \theta_2^2) \theta_1 \theta_2 \\
4 W'' (\theta_1^2 - \theta_2^2) \theta_1 \theta_2 &
2 W + 4 W' \theta^2 + 8 W'' \theta_1^2 \theta_2^2
\end{array}
\right] f(\vectheta)
\end{equation}
and with
\begin{equation}	\label{eq:esdef}
e^s_\alpha \equiv  {1\over T} \int d^2 \theta
\left[
\begin{array}{c}
\theta_1^2 - \theta_2^2 \\
2 \theta_1 \theta_2
\end{array}
\right] (6 W' + 2 W'' \theta^2) f(\vectheta)
\end{equation}

We can use (\ref{eq:Psmear}) as follows:
{}From the observed stellar $e$'s,  --- which, being circular,
will have diagonal $P^s_{\alpha\beta}$  --- we can
infer $p_\alpha$ (as a function of position on the chip if necessary).
Having fit the stellar ellipticities to some reasonable
model, we can then use (\ref{eq:Psmear}) to calculate the necessary linear
correction
to the galaxy ellipticity. This restores the polarisation values to
what they would have been for an observer with a perfectly circular psf.

One could make some simplifications here:
the expectation values of the
off-diagonal terms in $X^s_{\alpha\beta}$ vanish for
randomly oriented objects, as does the difference between diagonal
terms, and we have, for the shift in the mean polarisation
\begin{equation}
\langle \delta e_\alpha \rangle = P^s p_\alpha
\end{equation}
with
\begin{equation}
P^s = {
\int d^2 \theta (2W + 4 W' \theta^2 + W'' \theta^4) f(\vectheta)
\over
\int d^2 \theta W \theta^2 f(\vectheta)} -
\langle e_\alpha e^s_\alpha \rangle / 2
\end{equation}
Thus if we simply use this reduced scalar polarizability
and apply the correction
$\delta e_\alpha = - P^s p_\alpha$ then we make no error {\sl on average\/},
though with a finite number of galaxies the precision of the
correction should be better if we use the full polarizability
tensor, and this is what we do.

The smear polarizability depends on both the weight function and
on the galaxy shape. With our detection and analysis
scheme the weight function scale length is derived from the actual
image shape, and one can then see on dimensional grounds
from from equations \ref{eq:Xsdef},\ref{eq:esdef} and from the
definition of the trace $T$ that
$P^s_{\alpha\beta}$ scales inversely as the area of the image.

\section{Shear Polarizability} 		\label{sec:shear}

In this section we attempt to calculate how the polarization values
change under the influence of a small, coherent gravitational shear.
This is superficially very similar to the calculation of appendix
\ref{sec:shear} and the goal is to calculate some kind of
`shear-polarizability' which provides a calibration factor to
convert raw polarization values to shear estimates. As before, the
result depends on the details of the weighting function $W(\theta)$
and on the shapes of the galaxy, but the situation is much more
difficult here if the galaxies are only poorly resolved, since
the polarizability depends on the shapes of the galaxies
before seeing which is not measurable.  Here we will calculate
the shear polarizability in the limit that seeing can be neglected.
For ground based images this will not be a very good approximation,
but it provides a starting point.

The perturbation to the surface brightness pattern due to gravitational lensing
is
\begin{equation} 		\label{eq:fsheared}
f'(\theta_i) = f(\theta_i - \phi_{,ij} \theta_j)
\end{equation}
where perturbed angles are understood to be measured relative to
the perturbed centroid position and where $\phi$ is the surface potential.

If we take $\phi_{,ij}$ in \ref{eq:fsheared} to be small, and
effectively constant over the size of the background galaxy
then on substituting in \ref{eq:Qijdefinition} we have
\begin{equation}
Q'_{ij} = Q_{ij} - \phi_{,lm} \int d^2 W(\theta) \theta_i \theta_j \theta_m
\partial f(\vec\theta) / \partial \theta_l
\end{equation}
or, on integrating by parts,
\begin{equation}
Q'_{ij} = Q_{ij} + \phi_{,lm} Z_{lmij}
\end{equation}
with
\begin{equation}
Z_{lmij} =  \int d^2 \theta z_{lmij} f(\vec\theta)
\end{equation}
where
\begin{equation}
z_{lmij} = {\partial (W(\theta) \theta_i \theta_j \theta_m) \over
\partial \theta_l}
= W(\delta_{il} \theta_j \theta_m + \delta_{jl} \theta_i \theta_m +
\delta_{ml} \theta_i \theta_j ) + 2 W' \theta_l \theta_m \theta_i \theta_j
\end{equation}
and where, as before, prime denotes differentiation wrt $\theta^2$.

Now the perturbation to the polarization is, from \ref{eq:edefinition},
\begin{equation}
\delta e_\alpha = {\delta Q_\alpha \over T} - e_\alpha{\delta T \over T}
\end{equation}
Writing
\begin{equation}
\phi_{,ij} =
\left[
\begin{array}{cc}
\kappa + \gamma_1 &  \gamma_2 \\
 \gamma_2 & \kappa - \gamma_1
\end{array}
\right]
\end{equation}
with $\kappa$ the dimensionless surface density and
$\gamma$ the shear,
we find
\begin{equation}
\delta Q_{ij} = (\kappa + \gamma_1) Z_{11ij} + (\kappa - \gamma_1) Z_{22ij}
 + 2 \gamma_2 Z_{12ij}
\end{equation}
Calculating the 1st order change in $e_\alpha$ using equation
\ref{eq:edefinition} we find a large
number of terms. Some of these involve the surface density
$\kappa$, which might seem
surprising.  These terms arise from the first order perturbation
to the trace $T$.  However, the expectation value for these terms
taken over the randomly oriented (to zeroth order) background
galaxies vanishes, and we therefore ignore them without fear of
introducing a bias.  The remaining terms are linear in $\gamma_\alpha$:
\begin{equation} \label{eq:Pshear}
\delta e_\alpha = P^\gamma_{\alpha\beta} \gamma_\beta
\end{equation}
with
\begin{equation}
P^\gamma_{\alpha\beta} = X^\gamma_{\alpha\beta} - e_\alpha e^\gamma_\beta
\end{equation}
where $P^\gamma_{\alpha\beta}$ defines the {\sl shear polarizability\/} which
is quite
analogous to the smear polarisability defined in section \ref{sec:smear},
but now with somewhat different moments
\begin{equation}
X^\gamma_{\alpha\beta}= {1\over T}\int d^2 \theta
\left[
\begin{array}{cc}
2 W \theta^2 + 2 W'(\theta_1^2 - \theta_2^2)^2  &
4 W' (\theta_1^2 - \theta_2^2) \theta_1 \theta_2 \\
4 W' (\theta_1^2 - \theta_2^2) \theta_1 \theta_2 &
2 W \theta^2 + 8 W' \theta_1^2 \theta_2^2
\end{array}
\right] f(\vectheta)
\end{equation}
and
\begin{equation}
e^\gamma_\alpha \equiv 4 e_\alpha +  {2\over T} \int d^2 \theta
\left[
\begin{array}{c}
\theta_1^2 - \theta_2^2 \\
2 \theta_1 \theta_2
\end{array}
\right] \theta^2 W' f(\vectheta)
\end{equation}
Note that while the smear polarizability scales inversely as the area of the
object, the shear polarizability is a function only of the shape of
the image.

As with the smear polarizability, one can calculate a scalar
polarizability which gives the shift in the mean polarization
for intrinsically randomly oriented objects.
If we average over an ensemble of galaxies with the same
shape, but random orientations we find
\begin{equation}
\langle e_\alpha \rangle = P^\gamma \gamma_\alpha
\end{equation}
where
\begin{equation}
P^\gamma = 2 + { \int d^2 \theta W' \theta^4 f(\vectheta) \over
\int d^2 \theta W \theta^2 f(\vectheta)}
- \langle e_\alpha e^\gamma_\alpha \rangle
\end{equation}
so a fair estimate of the shear $\gamma_\alpha$ is given by
taking the mean of $e_\alpha / P^\gamma$.  However, as
with the smear polarizability, there is more information in the full
polarizability tensor, so this is what we use.


\section{Artificial Sources of Anisotropy} \label{sec:sources}

Weak lensing allows the possibility of measuring mass
fluctuations on large scales, but only if the precision can
be kept close to the minimum uncertainty implied by  the
statistically random intrinsic background galaxies.
It is therefore useful to consider what
systematic effects might arise in order than these can be
minimised or corrected for.

Many of the effects consist of a unidirectional smearing
of the images; in the absence of seeing each image
would then be a line of half-length $\delta \theta$, and the
psf anisotropy parameter as defined in equation \ref{eq:psfanisotropydef} is
then
$p_\alpha = \int d^2 \theta g \theta^2 /
\int d^2 \theta g = \delta \theta^2 / 3$.
The size of the effect on our shear estimates depends
on the size of the galaxy images.  For the degraded
HST data used here (smeared to $\simeq 0.5''$ seeing and with
$0.2''$ pixels) we find a median smear-polarizability
$P^m \sim 0.45 ({\rm pixels})^{-2}$, so the typical induced
polarization is
\begin{equation}
\delta e_\alpha \simeq 3.75 (\delta \theta / 1'')^2
\end{equation}
so smearing with a box-car psf of half width $0.1''$ say
would induce a polarization of $\simeq 4\%$.  In comparison,
the expected signal from large-scale structure is at about
the $\sim 1\%$
level.  With an idea then of what size of effect might be tolerated
we will now survey some of the obvious sources of psf anisotropy.

\subsection{Atmospheric Refraction}

Refraction by the atmosphere shifts images according to
\begin{equation}
z_a = z_t - R(z_a, \lambda)
\end{equation}
where $z_a,\ z_t$ are the apparent and real zenith distances
and the deflection angle $R(z, \lambda)$ is tabulated in
Allen (19??).
For reasonably small zenith angles
\begin{equation}
R \simeq R_0(\lambda) \tan z
\end{equation}
where $R_0\simeq 60''$ at sea level and at optical wavelengths,
falling to $R_0\simeq 35''$ at a 4000m site such as Mauna Kea.
The deflection angle is a fairly weak function of wavelength:
$d\log R / d\log \lambda \simeq \{7.9\%,4.5\%,3.2\%,1.9\%\}$
at wavelengths $\{4500,5500, 7000, 9000\}$ angstrom
corresponding to B,V,R and I wavebands.

The {\sl stretching\/} of images due to the atmospheric refraction
gives an apparent shear
\begin{equation}
2 \gamma = dR / d z_a \simeq R_0 z_a^2 / 6 \simeq 3 \times 10^{-5}
z_a^2
\end{equation}
which is entirely negligible for reasonable zenith angles
for the signals of interest. In fact
the shear rises to only about 1\% at $\sim 5\deg$ above the horizon.

There is however the possibility for this effect to introduce
a systematic anisotropic smearing in systems where the
guide star lies at a considerable angle from the CCD (at
CFHT the angle is $\simeq 0.5-1\deg$).  In a very long integration
the position of objects on the CCD will then drift by an
angle $\sim 0.12'' \Delta(z^2)$. This effect can easily be avoided
by keeping the exposures reasonably short (which tends to be desirable
for other reasons).  In any case, the effect is to produce an
essentially achromatic smearing which is constant over the
frame which can then be removed using the technique described
in appendix \ref{sec:smear}.

\subsection{Atmospheric Dispersion}

Atmospheric dispersion is potentially a more worrying
problem.  The atmosphere, acting as a prism, will
disperse any continuum emission over an angle
\begin{equation}
\delta \theta \sim (\delta \lambda / \lambda) R d \ln R / d \ln \lambda
\end{equation}
where the first factor is the half-width of the filter.  For
I-band observations we find $\delta \theta \simeq 0.12'' z$ giving
a polarization shift
\begin{equation}
\delta e \sim 0.05 z_a^2
\end{equation}
This is quite a large effect.  More worrying, the effect depends
on the spectrum of the  object. We measure the psf anisotropy from stars
which are continuum objects, but then apply the correction to galaxies.
The worst case would be a galaxy dominated by a single
very strong emission line which would then be
undispersed, and our `correction' would then actually introduce a systematic
error.  For less extreme objects and for realistic filters the dependence of
image profiles on the
object spectra is rather weak;
the galaxies and the stars may have systematically different spectral
slopes, but to
1st order a change in the slope of the spectrum simply displaces the centroid
of the object without
changing the shape of the psf.
To change the shape of the image requires there to be a change
in the slope of the spectrum over the typically rather narrow
pass-band of the filter.  For normal spectra, any such effect is therefore
quadratic in the filter width and is quite small.
Note however, that these problems would be exacerbated if one
tried to stack images taken in different passbands.

A quantitative estimate of the strength of this effect for realistic
galaxy/stellar SED's remains to be done,
but as the
effect is quadratic in zenith angle and prevention is better
than cure, the obvious solution is to try to keep to
small zenith angle if at all possible. Another possibility
is to employ an atmospheric dispersion compensator.

\subsection{Guiding Error}

One unavoidable effect is guiding error.  This will typically
contain a systematic effect and a stochastic effect (from
wind-shake for example). Any such effects are however achromatic and
constant across the CCD and (provided they are not too large)
can readily be corrected for.

\subsection{Optical Aberration}

This effect is highly site dependent. Spot diagrams
for CFHT (kindly provided by Derrick Salmon at CFHT) show that
according to the design specifications of the wide field corrector
the aberration should be
quite small for the FOCAM detector --- a single chip at the centre of the
very wide ($50'$) corrected field of view --- and also for
devices such as MOCAM, so insofar as the aberration is achromatic
the linearised correction developed here should work.
Eventually, we
can expect that the whole of the $\sim 1$ square degree corrected field will be
tiled with
chips, and the aberration increases considerably towards the edge of the
corrected field so this may become more of an issue.  The spot
diagrams show the distortion of the images to be fairly
achromatic, but there is some color dependence, but as with atmospheric
dispersion more work is needed to quantify the effect.  Note that even if the
psf does
vary considerably for stars as opposed to galaxies there may be
some remedy.  If the effect is repeatable (e.g.~a simple function of
distance from the axis of the telescope) then it may be possible to directly
measure any systematic over- or under-correction and modify the
correction appropriately.  Another way to remove repeatable psf anisotropy
is to rotate the instrument by $\pi/2$ between exposures.  Even if the
effects are not repeatable, if they vary slowly across the field then
as one will have a very large number of stars it may prove
possible to develop a model for the psf as a function of colour.  Finally,
if all else fails, it may prove necessary to work with narrower band filters
though this entails some overhead.

One interesting effect we have come across at CFHT was a
reproducible linear gradient of ellipticity across the CCD
frame.  The cause of this is suspected to have been an asymmetric
aberration of the primary mirror. This is known to cause
a psf anisotropy which changes sign as the chip is moved
up and down during the focussing, and we may well have been
seeing a slight misalignment of the chip and focal plane coupling to this
and generating a gradient in image polarisation.

\subsection{Image Addition}

Even under perfect observing conditions, some psf anisotropy
will inevitably arise from stacking multiple images.
We have done this using only linear translations and with
integer pixel shifts, and this can
easily introduce spurious polarization at the few percent
level.  The effect is achromatic and slowly varying and should
therefore be correctable.  It is possible to apply
fractional pixel shifting with some interpolation scheme, but
this does not really avoid the problem.  An often quoted advantage
of `sinc-interpolation' is that is preserves the lack of
correlation in the photon counting noise.  It does not however,
preserve the shape of signals, and does introduce psf anisotropy.

This effect can be more dangerous if one tries to stack
images which are rotated with respect to each other.  One then
obtains a moire pattern and the induced polarization will then vary
periodically
with a high spatial frequency. With the limited number of
stars it may be difficult or impossible to map this pattern, and
the correction will then fail.

\end{document}